\documentclass[twocolumn,aps,prb,eqsecnum,showpacs,floatfix]{revtex4}
\usepackage{graphicx,bm,amsmath}
\usepackage{times}

\begin{document}

\title{
Peierls ground state and 
excitations in the electron-lattice correlated system (EDO-TTF)$_2$X
}
 
\author{M.\ Tsuchiizu and Y.\ Suzumura}
\affiliation{
Department of Physics, Nagoya University, Nagoya 464-8602, Japan
}

\date{May 20, 2008}

\begin{abstract}
We investigate the exotic Peierls state in the 
 one-dimensional organic compound  (EDO-TTF)$_2$X, 
wherein the Peierls transition is 
  accompanied by the bending of molecules and 
  also by a fourfold  periodic array of 
  charge disproportionation
  along the one-dimensional chain.
Such a Peierls state,
   wherein the interplay between
   the electron correlation and the electron-phonon 
    interaction takes an important role, 
    is examined based on an extended Peierls$-$Holstein$-$Hubbard model 
that includes
 the alternation of the elastic energies
for both the lattice distortion  and the molecular deformation.
The model reproduces the experimentally observed pattern of  
the charge disproportionation
 and 
 there exists a metastable state
     wherein the energy takes a local minimum
       with respect to the lattice distortion
  and/or molecular deformation.
Furthermore,
we investigate the excited states 
 for  both the Peierls ground state and the metastable state
  by considering the soliton formation of electrons.
 It is shown that the soliton excitation from the metastable state 
       costs energy that is much smaller than that of 
       the Peierls state, 
  where the former is followed only by the charge degree of freedom
 and the latter is followed by that of spin and charge.
Based on these results,
 we  discuss the exotic photoinduced phase
 found in (EDO-TTF)$_2$PF$_6$.
\end{abstract}


\maketitle

\section{Introduction}

The quasi-one-dimensional molecular compound,  (EDO-TTF)$_2$PF$_6$,
(Refs.\ \onlinecite{Chollet2005,Onda2005,Koshihara2006,Ota2002,Drozdova2004})
(where EDO-TTF denotes ethylenedioxytetrathiafulvalene)
 has been studied as one of the central topics  
 showing a photoinduced phase transition.
 \cite{Tokura2006}
The 1/4-filled (EDO-TTF)$_2$PF$_6$ system 
exhibits a Peierls insulating state,
which is accompanied by a
  charge disproportionation and the 
  bending of the EDO-TTF molecules.\cite{Ota2002}
 Several experiments indicate that this Peierls 
  transition comes
  from the cooperation effect of the
  molecular deformation, charge ordering, and anion ordering. 
 The charge disproportionation along the one-dimensional chain  exhibits  
     a fourfold periodicity  given by a periodic  array of [0, 1, 1, 0];
  i.e.,  the pattern is an alternation of two charge-rich sites 
       and two charge-poor  sites.  
 A remarkable feature of the insulating state is the large bending 
   of the  EDO-TTF molecules, which does exist
   even  for the neutral single molecule. 
The degree of bending 
    depends on the valence of the EDO-TTF molecule.
  Each molecule  in the high-temperature metallic phase
    has a valence of $+0.5$,
  while  the valences in the insulating state
  are estimated as $+0.9$ and $+0.1$ for hole-rich and hole-poor sites, 
  respectively. \cite{Drozdova2004}
   In the  insulating state, the degree of bending   
     is enhanced  at the electron-rich  site  and  
       is suppressed at the electron-poor site;  i.e., 
    the bending (flattening) of the molecule 
is observed in the electron-rich (-poor) site. 
However, the mechanism beyond the kinetic-energy gain 
 is required in order to 
 explain the bending and/or flattening of the molecules.

Recent experimental studies have focused on
  the phase transition induced by a weak
  laser pulse; i.e., the photoinduced phase transition,  and 
  suggest that (EDO-TTF)$_2$PF$_6$
   exhibits a gigantic photoresponse 
  in the low-temperature phase.\cite{Chollet2005}
In the photoinduced phase, 
  the lattice distortion and the molecular deformation are relaxed
  and the system shows a metallic behavior.  
In addition, it has been discussed that
  the photoinduced phase shows an exotic behavior, and 
  is different from that in the high-temperature metal phase.
  \cite{Onda2005}
From these experiments, it has been argued that 
  a new phase, 
  which is not possible as a ground state and is
  related to a metastable state, 
  is achieved by the photoexcitation.
Therefore it is of particular interest to investigate
  the mechanism of  a  metallic behavior  in  such a metastable state, which 
  may be related to the electronic correlation.

\begin{figure*}[t]
\begin{minipage}{9cm}
\includegraphics[width=4.1cm]{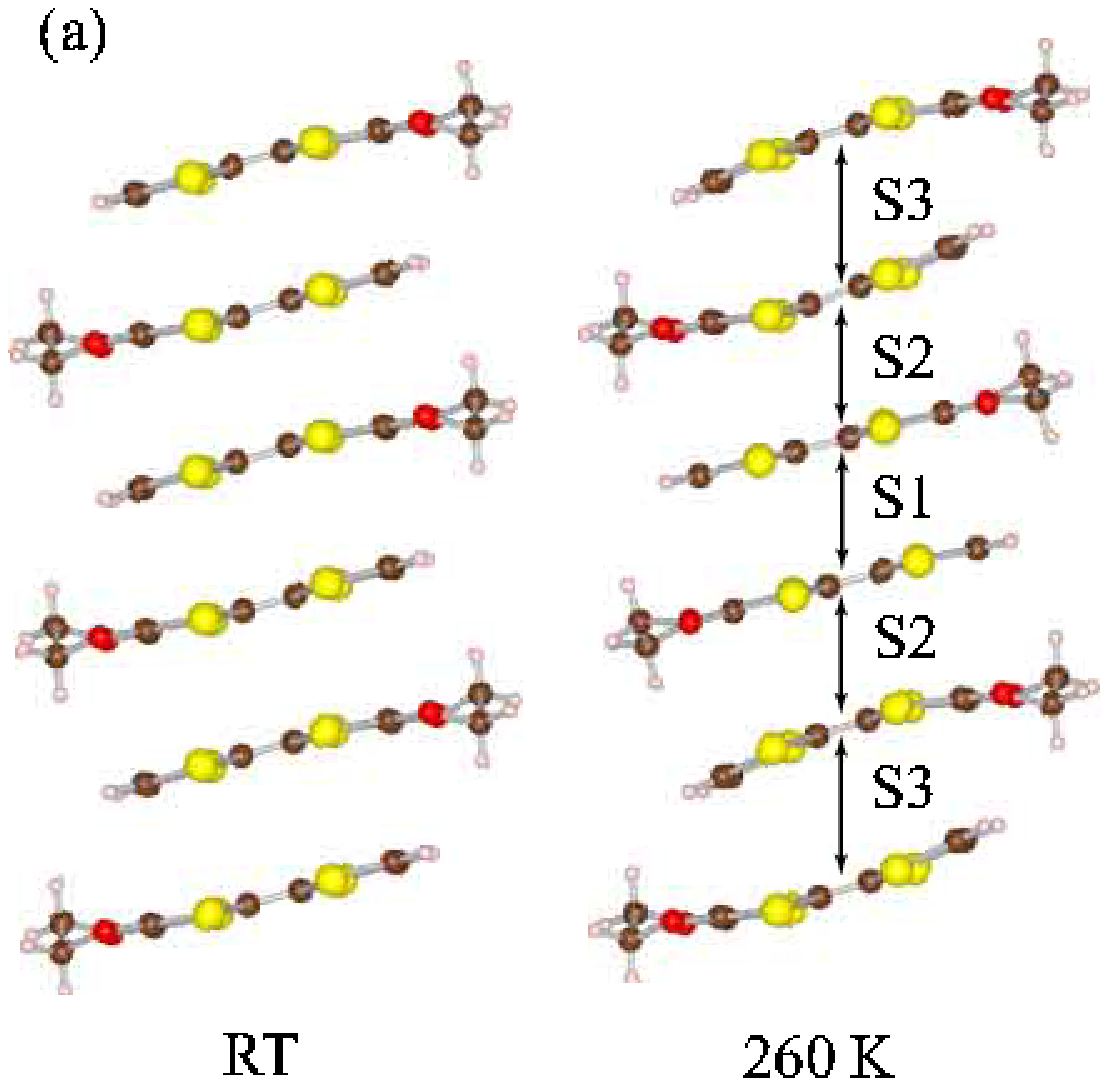}
\hspace*{3mm}
\includegraphics[width=4.1cm]{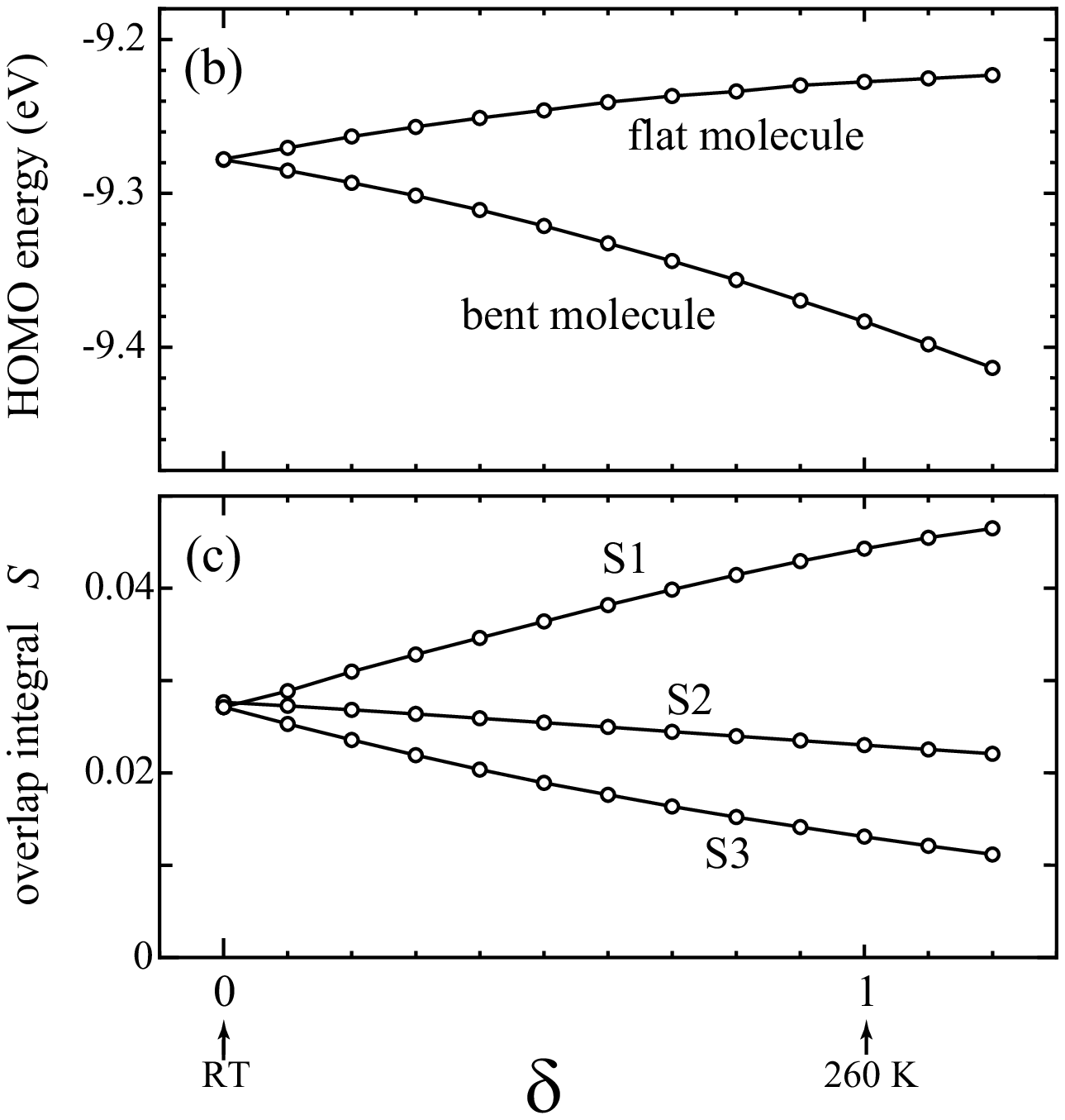}
\end{minipage}
\hspace*{1cm}
\begin{minipage}{7cm}
\caption{
(Color online)
(a) Crystal structure of (EDO-TTF)$_2$PF$_6$ at RT
  and 260 K. 
(b) The HOMO energy level of the EDO-TTF molecules as a function 
  of the distortion parameter $\delta$ (see text).
  $\delta=0$ corresponds to the case at RT and 
  $\delta=1$ to that at 260 K.
(c) The overlap integral as a function of $\delta$.
At RT, all molecules are equivalent and  S1$ = $S3, while S1$ \neq $S2. 
}
\label{fig:edo-ttf}
\end{minipage}
\end{figure*}

A theoretical investigation on 
the Peierls state in the one-dimensional (1D) quarter-filled system 
  has been performed
  based on the Peierls$-$Hubbard model including 
  the effect of the onsite Coulomb repulsion $U$
  and the intersite one $V$. \cite{Ung,Clay,Kuwabara2003,Seo2007JPSJ}
It has been clarified that
the spatial variation of the conventional charge-density wave 
  at the quarter-filled system is of the site-centered type;
  i.e., the charge density takes a maximum at a site.
On the other hand, in the Peierls state of (EDO-TTF)$_2$PF$_6$,
  it takes a maximum at the location between two neighboring sites;
  i.e., it is  of the bond-centered type.
It has been shown, within the mean-field theory, that 
  the experimentally observed pattern of charge disproportionation
  can be reproduced by taking into account 
 the alternation of the elastic energies.\cite{Omori2007}
It has also been shown that, due to the effect of the elastic-energy
modulation, the Peierls state and the charge-ordered (CO) state
  with a twofold periodicity compete with each other, and 
  there  exists the metastable state
     wherein the energy takes a local minimum
       with respect to the lattice distortion.
      \cite{Omori2007} 
The trigger of such a metastable state
 comes from 
 the competition between the electron correlation 
   and the electron-phonon interaction.
 The former favors the CO state with a twofold periodicity, while
  the latter supports the charge disproportionation 
  with a fourfold periodicity.
In the present paper, 
we study both the Peierls ground state and the metastable state,
     in terms of the phase representation based on bosonization.
Furthermore, we also examine the two kinds of excitations: 
(i) the excitation  followed by the lattice relaxation   and
 (ii) the purely electronic excitations.  
 In \S2, our model consisting of an electron-lattice system is  given 
  and a representation based on bosonization 
   is introduced. 
 The Peierls distortion with the bond order is taken  by considering  
  the alternation of the elastic constant for the lattice distortion. 
  In \S3, the ground state is analyzed 
    in terms of the phase variables  of spin and charge 
    to show the condition for the existence of the metastable state. 
     Using an extended Peierls$-$Holstein$-$Hubbard model 
       treated in terms of the bosonization method, 
       we demonstrate how the first-order transition occurs
        where the result is qualitative the same as the previous mean-field 
        result.\cite{Omori2007} 
 In \S4, the excitations from both the Peierls state and the metastable
 state  are examined by calculating the soliton formation energy. 
It is shown 
that   the latter is much  smaller than that of the former. 
Section V is devoted to  the summary and discussions.

\begin{figure*}[t]
\begin{minipage}{9cm}
\includegraphics[width=8.5cm]{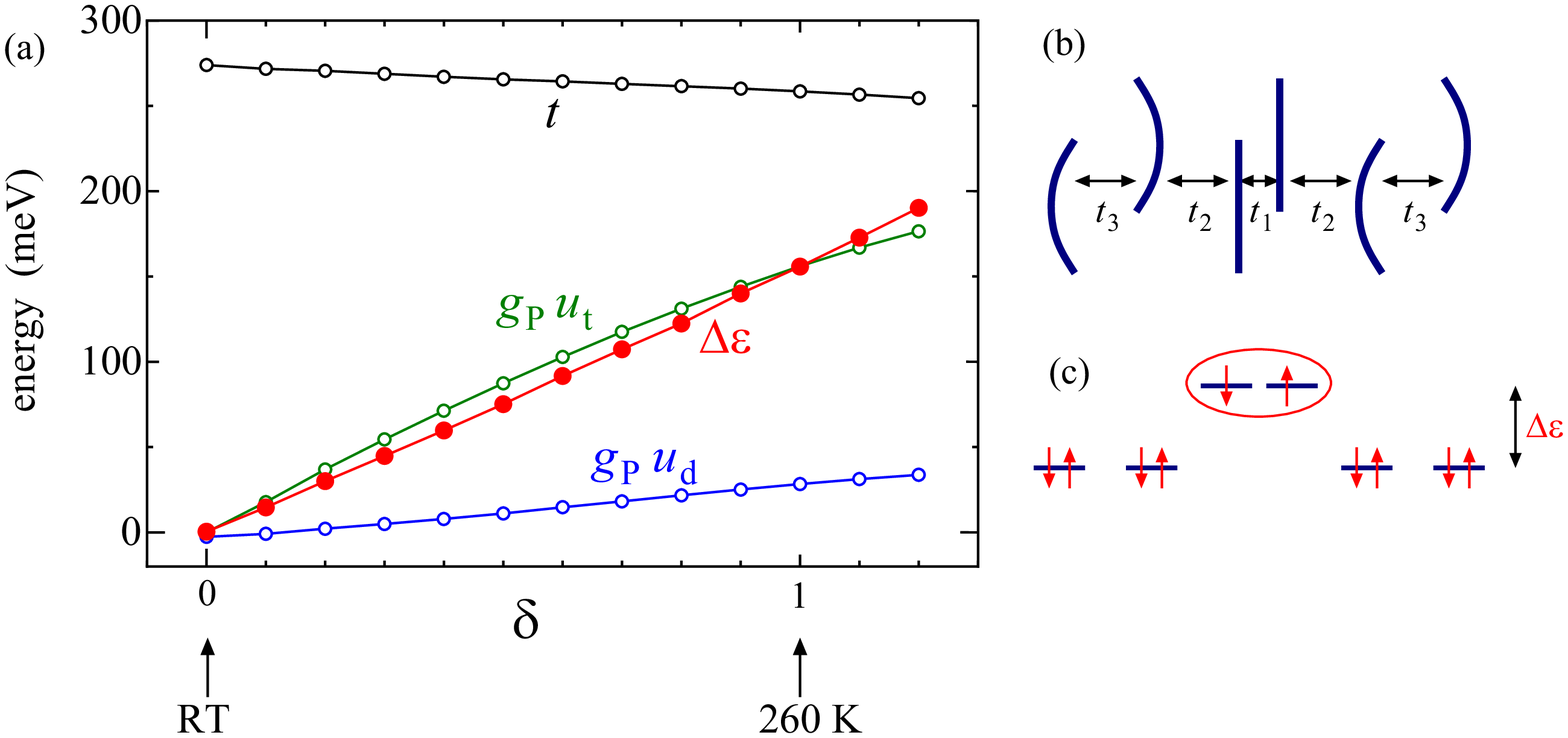}
\end{minipage}
\hspace*{1cm}
\begin{minipage}{7cm}
\caption{
(Color online)
(a)
 The HOMO energy difference $\Delta \varepsilon$ and the magnitude of 
  lattice tetramerization and dimerization  as a function 
  of the distortion parameter $\delta$.
(b) Schematic view of crystal structure
 of (EDO-TTF)$_2$PF$_6$ at 260 K and (c) 
the corresponding energy diagram,  
  where $\Delta\varepsilon = 156$ meV, 
  $t_1 = 443$ meV, $t_2 = 230$ meV, and $t_3= 131$ meV.
 }
\label{fig:edo-ttf2}
\end{minipage}
\end{figure*}


\section{Model for the EDO-TTF compound}

As a system exhibiting the correlated Peierls state, 
 we consider a tight-binding Hubbard model 
 coupled  with lattice distortion and molecular deformation, through 
    the electron-phonon interaction.
In this section, 
 by focusing on the Peierls state in (EDO-TTF)$_2$PF$_6$,
  we first examine the characteristic features of the 
  molecular deformation and the 
 lattice distortion of the EDO-TTF molecules, based on a 
  quantum-chemical calculation.
Next, we derive the model Hamiltonian of the 
  1D quarter-filled Peierls$-$Holstein$-$Hubbard model, and 
 we introduce the ``phases'' of the lattice distortion and of 
  the molecular deformation.

\subsection{Molecular deformation in (EDO-TTF)$_2$PF$_6$}

It has been confirmed that (EDO-TTF)$_2$PF$_6$ exhibits
 a first-order transition at around 280 K.
\cite{Ota2002}
The overlap integrals at room temperature (RT) and
   at 260 K have been estimated
 by an extended H\"uckel calculation\cite{Mori_extdh} based on 
an x-ray structure analysis.
At room temperature, 
  there is a very weak dimerization and
  the overlap integral is almost uniform along the stacking direction.
However,  at 260 K,
there is a strong variation among the overlap integrals, given by
  $S1$, $S2$, and $S3$,
along the stacking direction. \cite{Ota2002}

In order to make the situation clearer,
  we estimate the energy levels of the highest occupied molecular orbitals
 (HOMOs)  for the flat and bent molecules, and
 also estimate the overlap integrals, as a function of the
   lattice distortion and/or molecular deformation.
Here we simply perform the interpolation of the 
   coordinates for the respective atoms in the EDO-TTF molecules
between RT and 260 K by using  the relation
  $(x,y,z)=(x,y,z)_{\mathrm{RT}} + (\Delta x,\Delta y,\Delta z) \delta$,
where $\delta$ is the distortion parameter and
  $(\Delta x,\Delta y,\Delta z)
   \equiv (x,y,z)_{\mathrm{260 \,\, K}}-(x,y,z)_{\mathrm{RT}}$.
For both cases of RT and 260 K, 
the $y$ axis is chosen along the stacking direction [the vertical axis
  in Fig.\ \ref{fig:edo-ttf}(a)]
 and the $x$ axis is set to the projection of the 
  molecular-long direction to perpendicular to $y$
  [the horizontal axis in Fig.\ \ref{fig:edo-ttf}(a)].
At $\delta=0$ $(1)$, the atomic coordinates at RT (260 K) are 
  reproduced.
Using this relation and the extended H\"uckel method,
  \cite{Mori_extdh}
  we examine the $\delta$ dependence of 
  the HOMO energy levels [Fig.\ \ref{fig:edo-ttf}(b)] and 
the overlap integral [Fig.\ \ref{fig:edo-ttf}(c)].
Here we neglect the flipping disorder of the terminal ethylene group
  \cite{Ota2002}
  and use only the coordinates of the atoms with a high occupation rate.
We found that the HOMO energy level for the bent EDO-TTF molecule
  is sufficiently lowered due to the molecular deformation, and 
  that the overlap integrals do not change their sign as a function 
  of $\delta$.
We estimate the transfer integrals from the overlap integrals
  by using the empirical relation $t_i=S_i \times $ 10 eV.\cite{Mori_extdh}
At 260 K, the transfer integrals are estimated as
  $t_1= 442$ meV, 
  $t_2= 230$ meV, and 
  $t_3 = 131$ meV.
A notable feature of this compound is that the electron-rich EDO-TTF molecule 
  tends to bend and the hole-rich EDO-TTF molecule tends to be flat
 [Figs.\ 2(b) and 2(c)], \cite{Ota2002}
  where the valence of the former (latter) is 0.1 (+0.9).
The energy level of the HOMO is $\epsilon_{F}=-9.222$ eV for the flat
  EDO-TTF molecule, while
 $\epsilon_{B}=-9.442$ eV for the bent molecule.
This lattice distortion pattern shows
 a strong tetramerization $u_{t}$  in addition 
  to the lattice dimerization $u_{d}$, which are estimated from
\begin{subequations}
\begin{eqnarray}
t_1 &=& t+g_{P}u_{t} + g_{P}u_{d}, \\
t_2 &=& t - g_{P}u_{d}, \\
t_3 &=& t-g_{P}u_{t} + g_{P}u_{d}, 
\end{eqnarray}%
\label{eq:t_i}%
\end{subequations}
 where $g_{P}$ is the electron-phonon coupling constant.
The lattice distortion parameter $u_{t}$ 
  characterizes a modulation of fourfold periodicity; i.e., 
 the lattice tetramerization, and the
  parameter $u_{d}$ characterizes a modulation of
  twofold periodicity; i.e., the lattice dimerization.
It can be found that 
the strength of the lattice tetramerization is 
$g_{P}u_{t} = 156$ meV, 
while that of the lattice dimerization is
$g_{P}u_{d} = 29$ meV.

We note that the spatial variation of the lattice distortion
 corresponds to neither the conventional 
  $2k_F$ charge-density wave\cite{Ung,Clay} state nor
 the dimer-Mott+spin-Peierls (SP) state,\cite{Kuwabara2003,Seo2007JPSJ}
in which 
the lattice \textit{tetramerization} takes a maximum at $t_1$
(i.e., $g_{P}u_{t} > 0$), and 
the lattice \textit{dimerization} has a
maximum at $t_2$ (i.e., $g_{P}u_{d} < 0$).
In the dimer-Mott+SP state, which occurs in the presence of
 the lattice dimerization, 
  the system becomes effectively half filling and each electron is 
  localized at the location of each lattice dimer.
When the $2k_{F}$ Peierls instability is taken into account,
  such a paramagnetic insulator changes into the non-magnetic SP state.
 \cite{Kuwabara2003,Seo2007JPSJ}
Then, one finds $g_{P}u_{t} > 0$ and
 $g_{P}u_{d} < 0$ in the  dimer-Mott + SP state.
However, in the case of (EDO-TTF)$_2$PF$_6$, 
  both the lattice \textit{tetramerization} and 
  the lattice \textit{dimerization}
  take the maximum at $t_1$ (i.e., $g_{P}u_{t} > 0$ 
  and  $g_{P}u_{d} > 0$).
This lattice distortion pattern in (EDO-TTF)$_2$PF$_6$
  is qualitatively different from that in 
  the dimer-Mott + SP state and has not been discovered  in 
   previous theoretical studies.
In this sense, the Peierls ground state of (EDO-TTF)$_2$PF$_6$
 is quite exotic and, is examined, in the present paper, 
 by constructing a model that can reproduce such a ground state.

\subsection{Model Hamiltonian}

Based on the above consideration, 
 we introduce a 1D extended  Hubbard model 
 coupled  with the lattice through 
    the electron-phonon interaction. 
Since there are two molecules within the unit cell,
  we  introduce two sites referred to $l=1$ and $l=2$.
In this study, we are based on the hole picture; i.e., 
  there is one carrier per two sites.
Our Hamiltonian is given by 
\begin{eqnarray}
  H=H_{\mathrm{e}}+H_{\mathrm{e\mbox{-}ph}}
   +H_{\mathrm{ph}} , 
\label{eq:total_Hamittonian}
\end{eqnarray}
where the purely electronic part $H_{\mathrm{e}}$ is 
($j=1,\ldots,N$)
\begin{subequations}
\begin{eqnarray}
H_{\mathrm{e}} &=& 
-t \sum_{j,s} 
\left(c_{j,1,s}^\dagger c_{j,2,s}+\mathrm{H.c.} \right)
\nonumber \\ &&{}
-t \sum_{j,s} 
\left(c_{j,2,s}^\dagger c_{j+1,1,s}+\mathrm{H.c.} \right)
\nonumber \\ &&{}
+ U \sum_{j} 
 (n_{j,1,\uparrow} n_{j,1,\downarrow} 
+ n_{j,2,\uparrow} n_{j,2,\downarrow} )
\nonumber \\ &&{}
+ V \sum_{j} (n_{j,1} n_{j,2} + n_{j,2} n_{j+1,1})
-\mu \sum_{j,l(=1,2)} n_{j,l},
\nonumber \\
\label{eq:He}
\end{eqnarray}
 the electron-phonon coupling term $H_{\mathrm{e\mbox{-}ph}}$ is
\begin{eqnarray}
H_{\mathrm{e\mbox{-}ph}}
&=&
\sum_{j,s}
 g_{P} (u_{j,1}-u_{j,2})
\left(c_{j,1,s}^\dagger c_{j,2,s}+\mathrm{H.c.} \right)
\nonumber \\ && {}
+\sum_{j,s}
g_{P} (u_{j,2}-u_{j+1,1})
\left(c_{j,2,s}^\dagger c_{j+1,1,s}+\mathrm{H.c.} \right)
\nonumber \\ && {}
+\sum_{j,l(=1,2)} 
g_{\mathrm{H}}
v_{j,l} \left( n_{j,l} -\frac{1}{2}\right),
\label{eq:He-ph}
\end{eqnarray}
and the phonon term $H_{\mathrm{ph}}$ is
\begin{eqnarray}
H_{\mathrm{ph}}
&=&
 \frac{K_{P}}{2} \sum_{j} 
\left[
  (u_{j,1}- u_{j,2})^2
+ (u_{j,2}- u_{j+1,1})^2
\right]
\nonumber \\ && {}
+ \frac{K_{H}}{2} \sum_{j} (v_{j,1}^2 + v_{j,2}^2)
\nonumber \\ && {}
+ \sum_j F(u_{j,1},u_{j,2},v_{j,1},v_{j,2})
.
\label{eq:Hph}%
\end{eqnarray}%
\end{subequations}
The parameter $v_{j,l}$ represents the degree of molecular deformation
  wherein the shape of the molecule depends on  its sign; i.e., 
the molecule is bent for  $v_{j,l}>0$,  while
the molecule is flat for  $v_{j,l}<0$.
In Eq.\ (\ref{eq:He}),
 $c_{j,l,s}$ denotes the annihilation operator of an electron 
   with spin $s (=\uparrow, \downarrow)$ 
    at the $l$th site within the $j$th unit cell, and 
      $n_{j,l,s}=c_{j,l,s}^\dagger c_{j,l,s}^{}$,
      where $t$ is the transfer energy and  
        $\mu$ is the chemical potential. 
The parameters
    $U(>0)$ and $V(>0)$ denote the magnitudes 
    for the on-site and nearest neighbor interactions. 
In Eq.\ (\ref{eq:He-ph}),
$g_{H}$ is the electron-phonon coupling constant 
  of the on-site (Holstein) type, and
the Peierls-type electron-phonon coupling constant
  is given by
 $g_{P}$.
In Eq.\ (\ref{eq:Hph}), 
the parameters $K_{H}$ and $K_{P}$ are 
  the conventional elastic constant for the molecular deformation and
 the lattice distortion, respectively.
The quantity $F(u_{j,1},u_{j,2},v_{j,1},v_{j,2})$ represents
  the specific features in (EDO-TTF)$_2$PF$_6$, which will be
  defined in Sec.\ II C.

Here, we note that 
the present Hamiltonian can reasonably  account for the relation 
   between the valence of the molecule and 
  the molecular deformation $v_{j,l}$.
For simplicity, we consider the case of the single molecule, wherein
the Hamiltonian is given by
\begin{eqnarray}
H_{1} = g_{H} v \left( n - \frac{1}{2}\right)
+ \frac{1}{2} K_{H} v^2.
\end{eqnarray}
By minimizing $H_1$ with respect to $v$, we find that 
the lattice distortion depends on the charge on the molecule, e.g.,
\begin{eqnarray}
v =
\left\{
\begin{array}{ll}
\displaystyle
+\frac{g_{H}}{2K_{H}} & \mbox{for } n=0
\\ \\
\displaystyle
-\frac{g_{H}}{2K_{H}} & \mbox{for } n=1.
\end{array}
\right.
\end{eqnarray}
Thus, we obtain that the neutral molecule ($n=0$) becomes bent ($v<0$);
on the other hand,
  the ionic molecule ($n=1$) becomes flat ($v>0$).

\subsection{Modifications of phonon term $H_{\mathrm{ph}}$}

Here we examine the characteristic features of the crystal structure 
of (EDO-TTF)$_2$PF$_6$  [Fig.\ \ref{fig:edo-ttf}(a)],  
 for the model that reproduces the Peierls ground state. 
If $F(u_{j,1},u_{j,2},v_{j,1},v_{j,2})=0$ in  Eq.\ (\ref{eq:Hph}),
  the model reduces to the conventional Peierls$-$Holstein$-$Hubbard model
  at quarter filling.\cite{Ung,Clay,Kuwabara2003,Seo2007JPSJ}
Noting two kinds of molecules in the unit cell 
  at the high-temperature phase of the EDO-TTF compounds, 
  we introduce two kinds of elastic constants for the lattice
  distortion and the molecule deformation,
  within the unit-cell and between neighboring cells. 
Such a difference would play important roles on the low-energy properties.
We consider the term, $F(u_{j,1},u_{j,2},v_{j,1},v_{j,2})$, given by 
\begin{eqnarray}
&&
F(u_{j,1},u_{j,2},v_{j,1},v_{j,2})
\nonumber \\ && \qquad
=
- \frac{\delta K_{P}}{4} \sum_{j} 
\left[
  (u_{j,1}- u_{j,2})^2
- (u_{j,2}- u_{j+1,1})^2
\right]
\nonumber \\ && \qquad
-\frac{\delta K_{H}}{2} \sum_j
(v_{j,1} + v_{j,2})^2
,
\label{eq:Hph_mod}
\end{eqnarray}
where the $\delta K_{P}$ term comes from the difference between the 
  elastic constants  for the \textit{lattice distortion}, 
  within the unit cell and between  neighboring cells.
The $\delta K_{H}$ term 
  represents the difference in the energy gain 
  of the \textit{two-molecular deformation} within the unit cell.
The intrinsic property of the bending of the EDO-TTF molecules would play
  important roles for the lattice distortion, 
since a pair of two molecules faces each other in the unit cell. 
\cite{Ota2002}
In fact, in the ordered phase, 
one pair of molecules  shows strong bending 
and another pair of the molecules becomes more flat
[Fig.\ \ref{fig:edo-ttf}(a)].
Such an effect due to the pairing of the EDO-TTF molecules
 can be taken into account in our model 
  by considering the additional terms 
of Eq.\ (\ref{eq:Hph_mod}).

From the mean-field approach, \cite{Omori2007}
 it has been pointed out that the effect of $\delta K_{P}$ 
is important to determine the multistable states  
  induced by the lattice distortion.
The details of the effects of  
$\delta K_{P}$ and $\delta K_{H}$ 
are discussed  in Sec.\ III.

\subsection{Lattice distortion and molecular deformation}

In general, the lattice distortion $u_{j,l}$
 in  quarter-filled systems
can be decomposed into the 
  component with a twofold periodicity (i.e.,  dimerization),
 $u_{d}$, 
  and 
  that with a fourfold periodicity (i.e., tetramerization), $u_{t}$.
\cite{Clay,Kuwabara2003,Sugiura2003} 
Even for the freedom of the molecular deformation, $v_{j,l}$,
  it can be decomposed into the component with a
  twofold periodicity, $v_{d}$, and that with 
  a fourfold periodicity, $v_{t}$.
We introduce these quantities by using the relations:
\begin{subequations}
\begin{eqnarray}
u_{j,1}
&=&
+(-1)^j
\frac{u_{t}}{\sqrt{2}} \cos\xi - \frac{u_{d}}{2},
\\
u_{j,2}
&=&
-(-1)^j
\frac{u_{t}}{\sqrt{2}} \sin\xi + \frac{u_{d}}{2},
\\
v_{j,1}
&=&
+(-1)^j
\frac{v_{t}}{\sqrt{2}} \cos\zeta - \frac{v_{d}}{2},
\\
v_{j,2}
&=&
+(-1)^j
\frac{v_{t}}{\sqrt{2}} \sin\zeta + \frac{v_{d}}{2}.
\end{eqnarray}%
\label{eq:uj-vj}%
\end{subequations}
The phases $\xi$ and $\zeta$ determine the spatial pattern of the 
tetramerizations $u_{t}$ and $v_{t}$, respectively.

For the later convenience,
 we renumber the site index $j$ as
  $i=2j+(l-1)$; accordingly,
  we rewrite $c_{j,l,s}\to c_{i,s}$.
By using  Eq.\ (\ref{eq:uj-vj}),
  the Hamiltonian (\ref{eq:total_Hamittonian}) is
   rewritten as  
\begin{eqnarray}
H
&=& 
-  \sum_{i,s} 
\left[
  t- g_{P} u_{t} 
     \cos \left(\frac{\pi}{2} i +\xi - \frac{\pi}{4}\right)
 + (-1)^i g_{P} u_{d}
\right] 
\nonumber \\ &&{}
\qquad \times
\left(c_{i,s}^\dagger c_{i+1,s}+\mathrm{H.c.} \right)
\nonumber \\ &&{}
+
\sum_{i} 
g_{H}
\frac{v_{t}}{\sqrt{2}} 
 \cos \left(\frac{\pi}{2} i -\zeta \right)
  \left( n_{i} -\frac{1}{2}\right)
-\mu \sum_{i} n_{i}
\nonumber \\ &&{}
+ U \sum_{i} n_{i,\uparrow} n_{i,\downarrow} 
+ V \sum_{i} n_{i} n_{i+1}
\nonumber \\ &&{}
+ N \frac{K_{P}}{4} u_{t}^2 
- N \frac{\delta K_{P}}{4} u_{t}^2 \sin 2\xi
+ N \frac{K_{P}}{2} u_{d}^2 
\nonumber \\ &&{}
+ N \frac{K_{H}-\delta K_{H}}{4} v_{t}^2 
- N \frac{\delta K_{H}}{4} v_{t}^2 \sin 2\xi
+ N \frac{K_{H}}{4} v_{d}^2 .
\nonumber \\
\label{eq:Hi}
\end{eqnarray}

\begin{figure}[b]
\includegraphics[width=6.5cm]{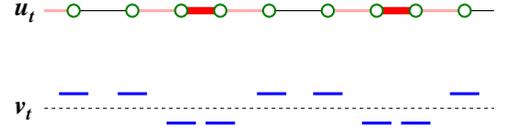}
\caption{
(Color online)
The modulation 
  pattern of the transfer integral  induced 
  by the lattice distortion $u_{t}$ and
 that of the onsite energy level 
  induced by the molecular deformation $v_{t}$.
In this view,  we have considered only the modulations of 
  tetramerization; i.e., of fourfold periodicity.
 }
\label{fig:model}
\end{figure}

Now, we specify the phases of $\xi$ and $\zeta$, which 
  reproduce the Peierls state observed in 
 (EDO-TTF)$_2$PF$_6$.
  The crystal structure of (EDO-TTF)$_2$PF$_6$ for the metallic state 
   at  high temperature 
  shows an alternation of the bending of the molecule,
 where all of the molecules are identical crystallographically.\cite{Ota2002}
  For the insulating state at low temperature,  
    the neutral molecule  exhibits a large bending 
   and the ionic molecule becomes rather flat,
     where  the adjacent two molecules at the neutral sites 
  are oppositely located with the same degree of bending
     and  becomes  convex outside.
The transfer integral between adjacent bent molecules
  becomes small while that between adjacent flat molecules 
  becomes large.
Such a pattern of the molecular deformation and 
 lattice distortion can be reproduced by 
  setting $\xi =\zeta =\pi/4$.
When $\xi =\zeta =\pi/4$,
  the transfer-energy modulation is given by 
\begin{eqnarray}
(t_3, t_2, t_1,t_2,t_3,t_2,t_1,t_2) ,
\label{eq:mod_u}
\end{eqnarray}
 for $i=0,1,2,3,4, \ldots$,
where $t_1$, $t_2$, and $t_3$ are the same as those given in 
  Eq.\ (\ref{eq:t_i}).
The schematic view of the modulation pattern is shown in Fig.\ 
  \ref{fig:model}.
  The site-energy modulation is given by
\begin{eqnarray}
\frac{1}{2} g_{H} 
   (+v_{t},+v_{t},-v_{t},-v_{t},
    +v_{t},+v_{t},-v_{t},-v_{t},
  \cdots),
\qquad
\label{eq:mod_v}
\end{eqnarray}
 for $i=0,1,2,3,4,\ldots$, 
  where the component of the molecular deformation 
  with the twofold periodicity $v_{d}$ 
  vanishes for (EDT-TTF)$_2$PF$_6$ 
(see also Fig.\ \ref{fig:model}).

\section{Bosonization Representation}

Here, we represent the Hamiltonian in terms of bosonic phase variables.
\cite{Giamarchi_book}
By setting $a$ as the lattice constant,
the electron density operator can be represented as
\cite{Giamarchi_book}
\begin{eqnarray}
\frac{n_i}{a}
&=&
\frac{1}{4a} +
\frac{1}{\pi}
  \frac{d\phi_\rho}{dx}
- \frac{2}{\pi a} \sin(2k_Fx+\phi_\rho)\cos \phi_\sigma
\nonumber \\ && {}
- \frac{2c}{\pi a} \cos (4k_Fx+2\phi_\rho) ,
\label{eq:phase_density}
\end{eqnarray}
where $x=ia$ and  $c$ is a nonuniversal constant. 
The Fermi momentum is $k_F=\pi/(4a)$.
The quantities $\phi_\rho$ and $\phi_\sigma$ are bosonic phase 
variables and  Eq.\ (\ref{eq:phase_density}) is justified 
 by the microscopic representation of the field operator.

Here we briefly recall the relation between the lattice distortion 
 and electron density modulation, based on the phase-variable representation.
We tentatively assume $c=0$.
If the phases are locked at 
  $(\langle \phi_\rho\rangle, \langle \phi_\sigma \rangle)=(\pi/4,0)$,
the expectation value of   the charge-density operator 
  are given by 
\begin{eqnarray}
\langle n_i \rangle
 &=&
\frac{1}{4}
- \frac{2}{\pi} \sin(2k_Fx +\pi/4) . 
\end{eqnarray}
The pattern of charge modulation 
  is given by 
$\langle n_j \rangle
   = (\, \circ \, \circ \, \bigcirc \, \bigcirc \, \circ \, \ldots \, )$
 for $j=0,1,2,3,4,\ldots$, which 
is compatible with the lattice distortion 
  and molecular deformation given in 
 Eqs.\ (\ref{eq:mod_u}) and (\ref{eq:mod_v}).
This means that 
the tetramerization pattern observed in 
  (EDO-TTF)$_2$PF$_6$ is reproduced 
  if the phases are locked at $\xi=\zeta =\pi/4$
  and 
$(\langle \phi_\rho\rangle, \langle \phi_\sigma \rangle)=(\pi/4,0)$.

Based on the phase-variable representation, 
 the bosonized  Hamiltonian
\cite{Fukuyama_review,Giamarchi_book,Tsuchiizu2001}
 is obtained  as 
$ H = \int dx \, \mathcal{H}$, where
\begin{widetext}
\begin{eqnarray}
\mathcal{H}
&=&
\frac{v_\rho}{4\pi}
\left[
 \frac{1}{K_\rho}(\partial_x \phi_\rho )^2
+ K_\rho(\partial_x \theta_\rho )^2
 \right]
+
\frac{v_\sigma}{4\pi}
\left[
 \frac{1}{K_\sigma}(\partial_x \phi_\sigma )^2
+ K_\rho(\partial_x \theta_\sigma )^2
 \right]
+ \frac{g_{1/4}}{2\pi^2 a^2}  \cos 4\phi_\rho
 + \frac{g_\sigma}{2\pi^2 a^2} \cos 2\phi_\sigma
\nonumber \\ && {} 
-\frac{2}{\pi a} g_{P}u_{t}
   \cos\left(\phi_\rho - \xi  \right)\cos \phi_\sigma
- \frac{1}{\sqrt{2}\pi a}
g_{H} v_{t}
 \sin \left(\phi_\rho +\zeta \right)
 \cos \phi_\sigma
+
\frac{U}{\pi^2 t a}  g_{P} u_{d}
\sin 2 \phi_\rho
+
 \frac{U} {2\sqrt{2}\pi^2 t a} 
g_{H} v_{d}
 \cos 2\phi_\rho
\nonumber \\ && {} 
+  \frac{K_{P}}{4a} u_{t}^2 
-  \frac{\delta K_{P}}{8a} u_{t}^2 \sin 2\xi
+  \frac{K_{P}}{2a} u_{d}^2 
+  \frac{K_{H}-\delta K_{H}}{8a} v_{t}^2 
-  \frac{\delta K_{H}}{8a} v_{t}^2 \sin 2\zeta
+  \frac{K_{H}}{8a} v_{d}^2 ,
\qquad
\label{eq:phase_Hamiltonian}
\end{eqnarray}
\end{widetext}
In Eq.\ (\ref{eq:phase_Hamiltonian}), 
$\theta_\nu$ ($\nu=\rho$ and $\sigma$) is the phase variable 
which is canonically conjugate to $\phi_\nu$ and  
$ [\phi_\nu(x),\theta_\nu(x')]
 = i \pi \mathrm{sgn} (x-x')$.
The parameters $v_\rho$ and $v_\sigma$ are the velocity of the
  charge and spin excitations, and 
$K_\rho$ and $K_\sigma$ are the Tomonaga$-$Luttinger parameters.
It is noted that $K_\rho<1$ for the repulsive interaction
  and $K_\sigma=1$ for the paramagnetic state.
 From the perturbative calculation, \cite{Tsuchiizu2001}
  the coupling $g_{1/4}$ is given by
  $g_{1/4}\propto U^2(U-4V)$, where it is noticed that 
     $g_{1/4}$ is positive for small $V$ but becomes negative
  for large $V$.
For small $V$, the phase $\phi_\rho$ is locked at 
$\langle \phi_\rho  \rangle=\pi/4$(mod $\pi/2$); while  
  for large $V$ the phase is locked at 
$\langle \phi_\rho  \rangle=0$(mod $\pi/2$).

In terms of the bosonized Hamiltonian, 
 we examine the effect of additional terms including 
 $\delta K_{P}$ and $\delta K_{H}$.
First, we discuss the state in the case of 
 $\delta K_{P}=\delta K_{H}=0$.
The locking position of the phases $\xi$ and $\zeta$ 
is determined by $\phi_\rho$; i.e.,
 $\xi = \phi_\rho$ and $\zeta = \pi/2 -\phi_\rho$.
Thus, the total Hamiltonian can be expressed as
\begin{eqnarray}
\mathcal{H}|_{\delta K=0}
&=&
\frac{v_\rho}{4\pi}
\left[
 \frac{1}{K_\rho}(\partial_x \phi_\rho )^2
+ K_\rho(\partial_x \theta_\rho )^2
 \right]
\nonumber \\ && {}
+
\frac{v_\sigma}{4\pi}
\left[
 \frac{1}{K_\sigma}(\partial_x \phi_\sigma )^2
+ K_\rho(\partial_x \theta_\sigma )^2
 \right]
\nonumber \\  && {}
+ \frac{g_{1/4}}{2\pi^2 a^2}  \cos 4\phi_\rho
 + \frac{g_\sigma}{2\pi^2 a^2} \cos 2\phi_\sigma
\nonumber \\ && {} 
-\frac{2}{\pi a} g_{P}u_{t}
 \cos \phi_\sigma
- \frac{1}{\sqrt{2}\pi a}
g_{H} v_{t}
 \cos \phi_\sigma
\nonumber \\ && {} 
+
\frac{U}{\pi^2 t a}  g_{P} u_{d}
\sin 2 \phi_\rho
+
 \frac{U} {2\sqrt{2}\pi^2 t a} 
g_{H} v_{d}
 \cos 2\phi_\rho
\nonumber \\ && {} 
+  \frac{K_{P}}{4a} u_{t}^2 
+  \frac{K_{P}}{2a} u_{d}^2 
+  \frac{K_{H}}{8a} v_{t}^2 
+  \frac{K_{H}}{8a} v_{d}^2 ,
\qquad
\label{eq:phase_Hamiltonian_nodeltaK}
\end{eqnarray}
which can reproduce the previous results
 (e.g., Fig.\ 4  in Ref.\ \onlinecite{Clay}) qualitatively.
The undistorted state can be obtained 
for a small electron-phonon coupling
 of $g_{P}$ and $g_{H}$.  
By increasing $g_{P}$ only, 
we obtain a solution where  $u_{d}$ becomes finite  and
  the phases $\phi_\rho$ and $\xi$ are locked at 
$\phi_\rho=\xi=\pi/4$(mod $\pi$) (if $u_{d}<0$) 
or at
$\phi_\rho=\xi=3\pi/4$(mod $\pi$) (if $u_{d}>0$).
This state corresponds to the $4k_F$ bond-order wave (BOW) state
 in Ref.\ \onlinecite{Clay}.
By further increasing $g_{P}$, 
 we obtain a state wherein the tetramerizations
  $u_{t}$ and $v_{t}$ become finite.
For $u_{d}<0$, 
  the lattice distortion [Eq.\ (\ref{eq:uj-vj})] is given by
\begin{subequations}
\begin{eqnarray}
u_{j,1}
&=&
+ \frac{|u_{d}|}{2}
+(-1)^j
\frac{u_{t}}{2} ,
\\
u_{j,2}
&=&
 - \frac{|u_{d}|}{2}
-(-1)^j
\frac{u_{t}}{2} .
\end{eqnarray}%
\label{eq:uj}%
\end{subequations}
Then the resultant hopping integrals are given by
\begin{subequations}
\begin{eqnarray}
t_1
&=&
t- g_{P}  |u_{d}| 
 + g_{P} u_{t},
\\
t_2
&=&
t + g_{P} |u_{d}| ,
\\
t_3
&=&
t- g_{P}  |u_{d}|
 - g_{P} u_{t}.
\end{eqnarray}%
\label{eq:uj}%
\end{subequations}
where 
$t_1=[t-g_{P} (u_{1,1}-u_{1,2})]$,
 $t_2=[t-g_{P} (u_{1,2}-u_{2,1})]$, and
 $t_3=[t-g_{P} (u_{j+1,1}-u_{j+1,2})]$. 
We note that this state is nothing but 
  the bond-charge-density wave (BCDW) state in Ref.\ \onlinecite{Clay}
and the dimer-Mott+SP state in Ref.\ \onlinecite{Kuwabara2003}.

Next we examine the effect of $\delta K_{P}$ and
 $\delta K_{H}$
From Eq.\ (\ref{eq:phase_Hamiltonian}),
we can immediately find that, if 
$\delta K_{P} >0$, the phase $\xi$ tends to 
 be  locked at $\xi=\pi/4$(mod $\pi$).
In addition, the phase $\zeta$ also tends to be locked 
  at $\zeta=\pi/4$(mod $\pi$) due to $\delta K_{H}(>0)$.
From Eq.\ (\ref{eq:phase_Hamiltonian}), we can immediately find that
the most plausible set of the locking pattern of the phases 
  in the Peierls state is given by 
   $(\xi,\zeta,\phi_\rho,\phi_\sigma)=(\pi/4,\pi/4,\pi/4,0)$.
Thus, the corresponding  charge modulation favors the bond-centered type.
In the following analysis, we restrict ourselves to the case of
  $\xi=\zeta=\pi/4$.
Another effect of $\delta K_{P}$ and $\delta K_{H}$
  is to decrease the elastic energy of the tetramerizations 
  $u_{t}$ and   $v_{t}$; i.e., 
the tetramerization is favorable compared to 
 the dimerization $u_{d}$ and the alternation $v_{d}$.
Thus we discard $u_{d}$ and $v_{d}$ in the following
 analysis.
In this case, the total Hamiltonian can be simplified as
\begin{eqnarray}
\label{eq:classical_hamiltonian}
\mathcal{H}
&=&
\frac{v_\rho}{4\pi}
\left[
 \frac{1}{K_\rho}(\partial_x \phi_\rho )^2
+ K_\rho(\partial_x \theta_\rho )^2
 \right]
\nonumber \\ && {}
+
\frac{v_\sigma}{4\pi}
\left[
 \frac{1}{K_\sigma}(\partial_x \phi_\sigma )^2
+ K_\rho(\partial_x \theta_\sigma )^2
 \right]
\nonumber \\  && {}
+ \frac{1}{2\pi^2 a^2}
V(u_{t},\phi_\rho,\phi_\sigma) ,
\end{eqnarray}
 where the potential term is given by
\begin{eqnarray}
&&
V(u_{t},\phi_\rho,\phi_\sigma)
=
 g_{1/4} \cos 4\phi_\rho
 + g_\sigma \cos 2\phi_\sigma
\qquad\qquad
\nonumber \\  
&& {} \qquad\qquad
- u_{t}
   \cos\left(\phi_\rho - \frac{\pi}{4}\right)\cos \phi_\sigma
+ \frac{K_{\mathrm{ph}}}{2} u_{t}^2 
,\quad
\label{eq:potential_hamiltonian}
\end{eqnarray}
Here  we have rescaled 
  $(4\pi a) [ g_{P} u_{t}
  + g_{H} v_{t}/(2\sqrt{2}) ] 
 \to u_{t}$ and $K_{\mathrm{ph}}$ is evaluated from the
  variational  method as
$ K_{\mathrm{ph}} \equiv 
[ (\tilde{K}_{P} + \tilde{K}_{H})
 / (\tilde{K}_{P} \tilde{K}_{H}) ]/(32a)$
 where 
$\tilde{K}_{P} \equiv
  (K_P-\delta K_{P}/2)/g_{P}^2$ and
$\tilde{K}_{H} \equiv
 4 (K_H-2\delta K_{H})/g_{H}^2$.
Hereafter we perform the classical treatment
 for the qualitative understanding of the metastable state 
  at zero temperature.

\section{Peierls State versus Charge-Ordered State}

In this section,  the ground state and the metastable state are examined 
 by calculating the minimum energy as a function of $u_{t}$. 
 Since the phase variables  $\phi_\rho$, $\phi_\sigma$, and the 
  lattice distortion $u_{t}$ are 
  spatially uniform, 
we can focus  only on the potential term 
 $V(u_{t},\phi_\rho,\phi_\sigma)$ by discarding 
 the first and second terms  of 
  Eq.\ (\ref{eq:classical_hamiltonian}).
The $u_{t}$ dependence of the  ground-state energy is estimated 
 by using 
 the stationary conditions for the phase variables, which are written as
 ($\nu = \rho$ and  $\sigma$) 
\begin{eqnarray}
   \frac{\partial V(u_{t},\phi_\rho,\phi_\sigma)}
   { \partial \phi_\nu} =0 .
 \label{eq:variation}
 \end{eqnarray}

\begin{figure}[t]
\includegraphics[width=6.5cm]{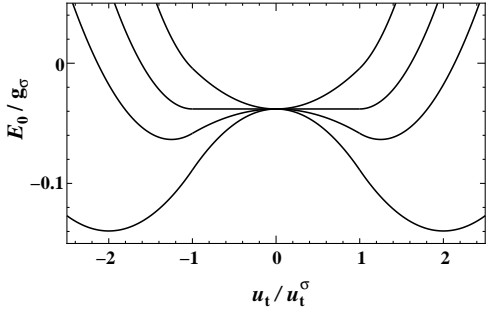}
\caption{
The  energy $E_0(u_{t})$ as a function of $u_{t}$, 
where $u_{t}^\sigma=4g_\sigma>0$, and we have set
 $g_{1/4}/g_\sigma=+1/4$.
The elastic constant is fixed as 
  $g_{\sigma}K_{\mathrm{ph}}=1/3$,
 $1/4$, $1/5$, and $1/8$, from top to bottom. 
The critical point is $g_{\sigma}K_{\mathrm{ph}}=1/4$.
}
\label{fig:potential-c}
\end{figure}

First we consider the case of $V=0$.
In this case, 
 the coupling constant of the commensurability energy 
  $g_{1/4}$ becomes positive, and then 
    the charge phase  $\phi_\rho$ is locked
        at $\phi_\rho^{\mathrm{opt}}=\pi/4$.
As for the spin phase $\phi_\sigma$, the optimized locking position is
   $\phi_\sigma^{\mathrm{opt}}
    =  \cos^{-1} \left(  u_{t}/ u_{t}^\sigma \right)$
        for $u_{t} < u_{t}^\sigma$, 
     and  $\phi_\sigma^{\mathrm{opt}} = 0 $
          for $u_{t} > u_{t}^\sigma $,
      where $u_{t}^\sigma \equiv 4g_\sigma $.
Then, the minimum energy 
  $E_0(u_{t})\equiv
 V_0(u_{t},\phi_\sigma^{\mathrm{opt}},\phi_\sigma^{\mathrm{opt}})
 /(2\pi^2a^2)$
 is given by 
\begin{widetext}
\begin{eqnarray}
 E_0(u_{t}) 
 =
\left\{
\begin{array}{ll}
\displaystyle
\frac{1}{2\pi^2 a^2}
\left[
  |g_{1/4}|
   - g_\sigma
   +  \left( \frac{ K_{\mathrm{ph}}}{2} - \frac{1}{8g_\sigma}
          \right) u_{t}^2 
\right]
&
\mbox{for } 0 < u_{t} < u_{t}^\sigma 
\\ \\
\displaystyle
\frac{1}{2\pi^2 a^2}
\left[
|g_{1/4}|
    + g_\sigma  - u_{t}
      + \frac{K_{\mathrm{ph}}}{2} u_{t}^2
\right]
&
\mbox{for } u_{t}^\sigma < u_{t}.
\end{array}
\right.
\label{eq:EforPierls}
\end{eqnarray}
\end{widetext}
From the coefficient of $u_{t}^2$ of Eq.\ (\ref{eq:EforPierls}), 
it is found that the state $u_{t}=0$ becomes unstable and 
  the Peierls state is obtained 
  for $g_\sigma  < 1/(4 K_{\mathrm{ph}})$.  
 On the other hand, for  $g_\sigma  > 1/(4K_{\mathrm{ph}})$, 
the undistorted state ($u_{t}=0$)
   is obtained, leading to the spin-density-wave (SDW) state.  
The typical $u_{t}$ dependences  of  $E_0(u_{t})$ 
  are shown in Fig.\ 
 \ref{fig:potential-c}.

\begin{figure}[t]
\includegraphics[width=6cm]{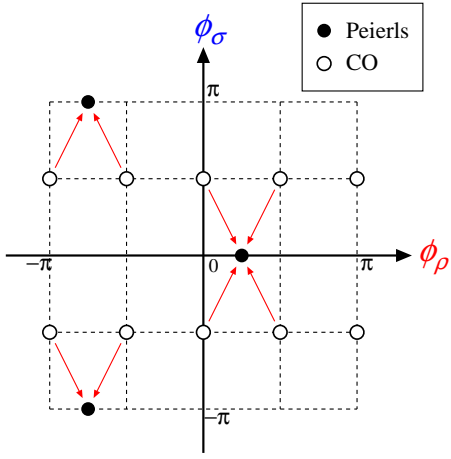}
\caption{
(Color online)
Positions of the locked phase fields $\phi_\rho$ and $\phi_\sigma$ 
  in the limits of the Peierls and CO states. 
The arrow indicates the schematic variation of the locking potential 
  with increasing $u_{t}$.
}
\label{fig:potmin}
\end{figure}

For the case of large $V$, 
 the coupling constant of the commensurability energy 
  $g_{1/4}$ changes its sign ($g_{1/4}<0$), and favors the
     locking position $\phi_\rho=0$(mod $\pi$),
  which is nothing but the CO state with a twofold periodicity; i.e., 
$\langle n_j \rangle
   = (\,  \bigcirc \, \circ \, \bigcirc \, \circ \, \cdots \, )$.
In this case, the region for    the lattice distortion 
  in divided by    two kinds of characteristic values 
   $u_{t}^\rho$ and    $u_{t}^\sigma$, which are given by
\begin{eqnarray}
u_{t}^\rho \equiv \sqrt{64|g_{1/4}|g_\sigma},
\quad
u_{t}^\sigma \equiv 4g_\sigma.
\end{eqnarray}
For small $u_{t}(<u_{t}^\rho,u_{t}^\sigma)$, 
  the phases $\phi_\rho$ and $\phi_\sigma$ 
       are locked at intermediate values, while 
 the phase $\phi_\nu$ takes 
 $ (\phi_\rho, \phi_{\sigma}) = (\pi/4,0)$, $(-3\pi/4, \pm\pi)$, $\ldots.$ 
 for large $u_{t}(>u_{t}^\rho,u_{t}^\sigma)$. 
For the intermediate value of $u_{t}$,  there are two possibilities:
  (i) $u_{t}^\rho \le u_{t}^\sigma$
  and (ii) $u_{t}^\sigma < u_{t}^\rho$. 
In the present paper, we focus on the former case, 
 $u_{t}^\rho \le u_{t}^\sigma$
(which is satisfied for  $g_{1/4}<0$ and $|g_{1/4}|<g_{\sigma}/4$),
since within the  mean-field calculation, \cite{Omori2007}
   only  the former case is realized and
  there would be some subtleties in the case of $|g_{1/4}|>g_{\sigma}/4$.

For case (i), 
the locking positions  for 
the phase variables $\phi_\rho$ and   $\phi_\sigma$ are
  analytically determined as 
\begin{equation}
\label{eq:rho_opt}
\phi_\rho^{\mathrm{opt}}
=
\left\{
\begin{array}{ll}
\displaystyle 
\frac{1}{2}
\sin^{-1} \left(
 \frac{u^2_{t}}{u^{\rho2}_{t}}
 \right)
&  \mbox{for }0<u_{t} < u_{t}^\rho
\\ \\
\displaystyle 
\frac{\pi}{4}
&  \mbox{for }  u_{t}^\rho < u_{t},
\end{array}
\right.
\end{equation}
and 
\begin{equation}
\phi_\sigma^{\mathrm{opt}}
=
 \left\{
\begin{array}{ll}
\displaystyle 
\cos^{-1}
\left[
  \frac{u_{t}}{u_{t}^\sigma}
\sqrt{
 \frac{1}{2}
\left(
  1+ \frac{u_{t}^2}{u^{\rho2}_{t}}
\right)
}
\right]   
& 
\mbox{for } 0<u_{t} < u_{t}^\rho
\\ \\
\displaystyle 
\cos^{-1}
\left(
  \frac{u_{t}}{u_{t}^\sigma}
  \right)
&
\mbox{for } u_{t}^\rho < u_{t}
              < u_{t}^\sigma 
\\ \\
0 
&
\mbox{for }u_{t}^\sigma < u_{t}.
\end{array}
\right.
\label{eq:sigma_opt}
\end{equation}
The schematic variations of  
$\phi_\rho^{\mathrm{opt}}$ and 
$\phi_\sigma^{\mathrm{opt}}$ as a function of $u_{t}$ 
are shown  in Fig.\ \ref{fig:potmin}.

\begin{figure}[t]
\includegraphics[width=6.5cm]{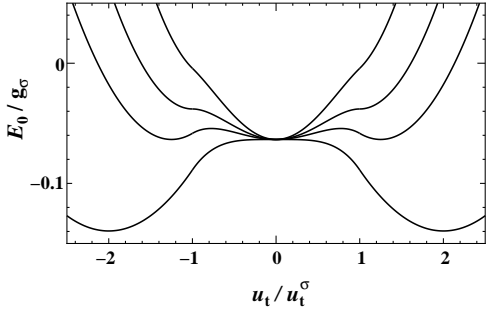}
\caption{
The energy $E_0(u_{t})$ as a function of $u_{t}$, 
where $u_{t}^\sigma=4g_\sigma$, and we have set
 $g_{1/4}/g_\sigma=-1/4$, 
The elastic constant is fixed as $g_{\sigma}K_{\mathrm{ph}}=1/3$,
 $1/4$, $1/5$, and $1/8$, from top to bottom. 
The transition point is $g_{\sigma}K_{\mathrm{ph}}=1/5$.
}
\label{fig:potential-b2}
\end{figure}

Now, we examine the energy as a function of $u_{t}$.
By inserting  Eqs.\ (\ref{eq:rho_opt}) and (\ref{eq:sigma_opt}) 
  into   Eq.\ (\ref{eq:potential_hamiltonian}),
the energy with fixed $u_{t}$ is given by
  $E_0(u_{t})\equiv
 V_0(u_{t},
     \phi_\rho^{\mathrm{opt}},\phi_\sigma^{\mathrm{opt}})
  /(2\pi^2a^2)$, 
  where
\begin{widetext}
\begin{eqnarray}
E_0(u_{t}) 
=
\left\{
\begin{array}{lll}
\displaystyle
\frac{1}{2\pi^2 a^2}
\left[
-|g_{1/4}|
- g_\sigma
+
\left(
  \frac{K_{\mathrm{ph}}}{2}
  - \frac{1}{16g_\sigma}
\right)
 u_{t}^2 
 - 
\frac{u^4_{t}}
{2^{11}|g_{1/4}| g_\sigma^2}
\right]
&\quad &
 \mbox{for  } 0< u_{t} < u_{t}^\rho 
\\
\\
\displaystyle
\frac{1}{2\pi^2 a^2}
\left[
+|g_{1/4}|
- g_\sigma
+  \left( \frac{K_{\mathrm{ph}}}{2} - \frac{1}{8g_\sigma}
   \right) u_{t}^2 
\right]
&&
 \mbox{for  }
    u_{t}^\rho < u_{t} < u_{t}^\sigma 
\\
\\
\displaystyle
\frac{1}{2\pi^2 a^2}
\left[
+|g_{1/4}|
+ g_\sigma
- u_{t}
+ \frac{K_{\mathrm{ph}}}{2}  u_{t}^2 
\right]
&&
 \mbox{for  } u_{t}^\sigma < u_{t}.
\end{array}
\right.
\label{eq:E0}
\end{eqnarray}
\end{widetext}
Typical $u_{t}$ dependencies of $E_0(u_{t})$
  are shown in Fig. \ref{fig:potential-b2}.
For $u_{t}>u_{t}^\rho$, 
  the energy $E_0(u_{t})$ takes a same form given 
  in Fig.\ \ref{fig:potential-c}.
It is noted that, in the present case,
  the energy around
  $u_{t}=0$ can become a local  minimum; i.e.,  
  the undistorted state is realized as a metastable state.
The actual ground state 
  is determined in order to 
  minimize  the energy $E_0(u_{t})$ with respect to
$u_{t}$. 
From Eq.\ (\ref{eq:E0}), we find that
the Peierls state with finite $u_{t}$ is obtained 
  in the case of 
\begin{eqnarray}
g_{\sigma}+|g_{1/4}| < \frac{1}{4 K_{\mathrm{ph}}} .
\end{eqnarray}
The amplitude of the lattice  distortion is given by
\begin{eqnarray}
u_{t}^0 =  \frac{1}{K_{\mathrm{ph}}} .
\end{eqnarray}
In addition, 
 {\it the condition for the Peierls state with 
  the metastable CO state ($u_{t}=0$)}  is given by
\begin{eqnarray}
g_\sigma + |g_{1/4}| < \frac{1}{4K_{\mathrm{ph}}}
 < 2 g_\sigma .
\end{eqnarray}
By further increasing  $K_{\mathrm{ph}} \{ >1/[4(g_\sigma +|g_{1/4}|)] \} $,
we find that 
  the energy at  $u_{t}=0$
   becomes lower than that 
  of finite $u_{t}$; i.e.,  
  the first-order phase transition from the Peierls state
   into the CO state occurs when  
  $1/(4K_{\mathrm{ph}}) =   g_{\sigma} + |g_{1/4}|$.

We note that the potential barriers from the Peierls state,
  $\Delta E_{\mathrm{P}} \equiv E_0(u_{t}^*)- E_0(u_{t}^0)$, 
  and that from the CO state,
  $\Delta E_{\mathrm{CO}} \equiv E_0(u_{t}^*)- E_0(0)$,
   are respectively given by 
\begin{subequations}
\begin{eqnarray}
\label{eq:height_CO}
\Delta E_{\mathrm{P}}
&=&
\frac{1}{2\pi^2 a^2}
\Biggl[
-2 |g_{1/4}| - 2 g_\sigma
+ \frac{1}{2K_{\mathrm{ph}}}
\nonumber \\ && {}
+
2^{9}  |g_{1/4}| g_\sigma^2
\left(
  \frac{K_{\mathrm{ph}}}{2}
  - \frac{1}{16 g_{\sigma}}
\right)^2
\Biggr]
, 
\\
\Delta E_{\mathrm{CO}}
&=&
\frac{1}{2\pi^2 a^2}
2^{9}  |g_{1/4}| g_\sigma^2
\left(
  \frac{K_{\mathrm{ph}}}{2}
  - \frac{1}{16 g_{\sigma}}
\right)^2 
, 
\qquad
\label{eq:height_Peierls}
\end{eqnarray}
\end{subequations}
 where 
$ u_{t}^* 
   \equiv 
  2^5 |g_{1/4}|^{1/2}g_{\sigma} \sqrt{K_{\mathrm{ph}}/2-1/(16g_{\sigma})}$
is the point at which the energy takes a local maximum.

\begin{figure}[t]
\includegraphics[width=8cm]{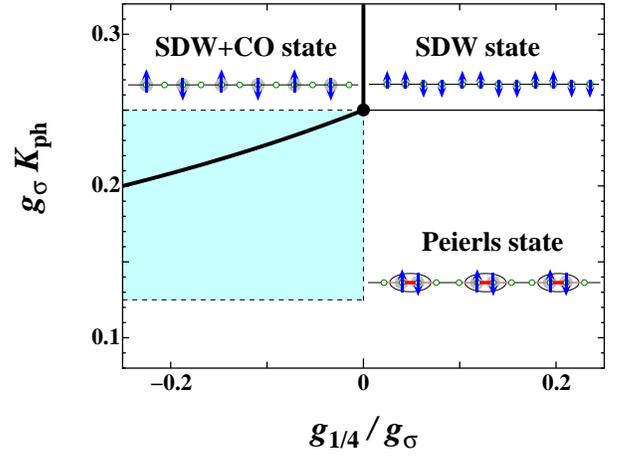}
\caption{
(color online).
Ground-state phase diagram on the plane of $g_{1/4}/g_\sigma$ and 
$g_\sigma K_{\mathrm{ph}}$.
In the shaded region, there  appears a metastable state.
}
\label{fig:phase_diagram}
\end{figure}

Based on the calculation of the ground state energy, 
 we obtain the phase diagram on the plane of $ g_{1/4}$ and $K_{\rm ph}$
  shown in Fig.\ \ref{fig:phase_diagram}.
For $g_{1/4} > 0$ in which the CO state is absent, 
   the pure SDW state is obtained for large $K_{\rm ph}$ 
 and  the Peierls state is obtained  for small $K_{\rm ph}$. 
 When  $g_{1/4} < 0$, 
 the Peierls state is obtained for 
  $4 K_{\rm ph} < 1/( g_{\sigma} + |g_{1/4}|)$ 
(region I)
 and 
the CO state coexisting with the SDW state is obtained for 
 $4 K_{\rm ph} > 1/( g_{\sigma} + |g_{1/4}|)$
 (region II).
The metastable state at $u_{t}=0$ is obtained 
  for $K_{\rm ph} > 1/(8 g_{\sigma})$ in region I,
 while 
 the metastable state at $u_{t} \neq 0$ is obtained 
  for
  $K_{\mathrm{ph}}<1/(4g_\sigma)$ in region II.
By taking into account the relations
  $g_{1/4}\propto U^2(U-4V)$ and  $g_\sigma \propto U$,
  \cite{Tsuchiizu2001} we can obtain  
 the ground-state phase diagram on the plane of $V$ and the
  elastic constant, 
  which reproduces that of the mean-field theory. \cite{Omori2007}
In section V, 
we examine the excited state in region I
 where the metastable state exists at $u_{t}=0$, 
 as shown in Fig.\ \ref{fig:potential-b2}.

\section{Excitations by Soliton Formations}

In this section,
we examine the excitations due to the soliton formations.
First, we derive the equations for spatially dependent phases, 
 which describe the variation  from those  of 
  the Peierls ground state and the metastable state around 
$u_{t} =0$.  
As shown in section IV, 
  the phase variables are uniform in space in the ground state. 
For simplicity, we do not consider the effect of 
 the quantum fluctuations induced by  $\theta_\rho$ and $\theta_\sigma$.
Since such effects give rise to a reduction in the amplitude 
  of the potential $V(u_{t},\phi_\rho,\phi_\sigma)$, 
 it may be qualitatively understood by introducing  the effective 
  coupling constants given by  
  $g_{1/4}^{\mathrm{eff}} = g_{1/4} \exp [- 8\langle \phi_\rho^2 \rangle]$,
  $g_{\sigma}^{\mathrm{eff}} 
     = g_{\sigma} \exp[- 2\langle\phi_\sigma^2\rangle] $,
   and 
   $u_{t}^{\mathrm{eff}}
       = u_{\mathrm{eff}} \exp[- \frac{1}{2}(\langle\phi_\rho^2\rangle
     + \langle \phi_\sigma^2 \rangle )]$,
  which can be estimated using the renormalization group treatment.
  \cite{Sugiura2003}

We examine  the classical Hamiltonian 
   given by $H_{\mathrm{cl}} = \int d x \, 
 \mathcal{H}_{\mathrm{cl}}$, where 
\begin{eqnarray}
\mathcal{H}_{\mathrm{cl}}
&=&
\frac{v_\rho}{4\pi K_\rho}
\left( \frac{d \phi_\rho(x)}{dx} \right)^2
+
\frac{v_\sigma}{4\pi K_\sigma}
\left( \frac{d \phi_\sigma(x)}{dx} \right)^2 
\nonumber \\  
&& {}
+\frac{1}{2\pi^2a^2}
V\biglb(u_{t}(x),\phi_\rho(x),\phi_\sigma(x) \bigrb)
.
\label{eq:classical_phase_hamiltonian}
\end{eqnarray}
Minimizing  Eq.\ (\ref{eq:classical_phase_hamiltonian})
 with respect to $u_{t}(x)$,
  $\phi_\rho(x)$, and $\phi_\sigma(x)$,\cite{Hara1983} 
  we obtain the following equations:
\begin{subequations}
\begin{eqnarray}
 u_{t}(x)
&=&
\frac{1}{K_{\mathrm{ph}}}
   \cos \left(\phi_\rho(x) - \frac{\pi}{4}\right)\cos \phi_\sigma(x) ,
\label{eq:u_opt}
   \\ 
0 &=& \frac{\pi v_\rho a^2 }{K_\rho} 
\frac{\partial^2 \phi_\rho(x)}{\partial x^2}
+4g_{1/4}\,  \sin 4\phi_\rho(x)
\nonumber \\ && {}
-  u_{t}(x) \sin \left(\phi_\rho(x) - \frac{\pi}{4}\right)
   \cos \phi_\sigma(x),
\label{eq:phi_rho_opt}
\\
0 &=&
\frac{\pi v_\sigma a^2 }{K_\sigma} 
 \frac{\partial^2 \phi_\sigma(x)}{\partial x^2}
+2 g_\sigma \,  \sin 2\phi_\sigma(x)
\nonumber \\ && {}
-  u_{t}(x) \sin \left(\phi_\rho(x) - \frac{\pi}{4}\right)
   \sin \phi_\sigma(x),   \qquad
\label{eq:phi_sigma_opt}
\end{eqnarray}%
\label{eq:soliton}%
\end{subequations}
which are self-consistently determined.
Note that substituting Eq.\ (\ref{eq:u_opt}) into Eqs.\ (\ref{eq:phi_rho_opt})
 and (\ref{eq:phi_sigma_opt}), we obtain the equations for 
 the phase variables $\phi_\rho(x)$  and $\phi_\sigma(x)$.

In the conventional picture of the soliton excitations 
 from the Peierls ground state,
  \cite{Takayama1980}
the charge and/or spin solitons are accompanied with the 
  soliton of lattice distortion 
 which connects the two minima of the potential.
Due to this effect, 
 the midgap state appears in the presence of the lattice distortion.
In the present case, however, 
  we have another excitation
  due to \textit{the lattice relaxation}.
As seen in Sec.\ IV,
such an  undistorted state ($u_{t}=0$)
  can be locally stabilized due to the competition between 
  the Peierls state and the CO state.  
In this section, first we verify that such a metastable state
  can be stabilized within the reasonable choice of parameters.
Next, we also consider the 
  purely electronic excitations by considering the 
  antiadiabatic limit where 
  the lattice distortion is assumed to be uniform, and 
we estimate the magnitude of the excitation gap.
These states are calculated to comprehend  the state relevant to
  reflectivity measurements in the Peierls phase and in the 
 photoinduced phase in (EDO-TTF)$_2$PF$_6$.

\subsection{Lattice relaxations: 
  The appearance of the charge-ordered domain in the Peierls state}

\begin{figure}[t]
\includegraphics[width=7cm]{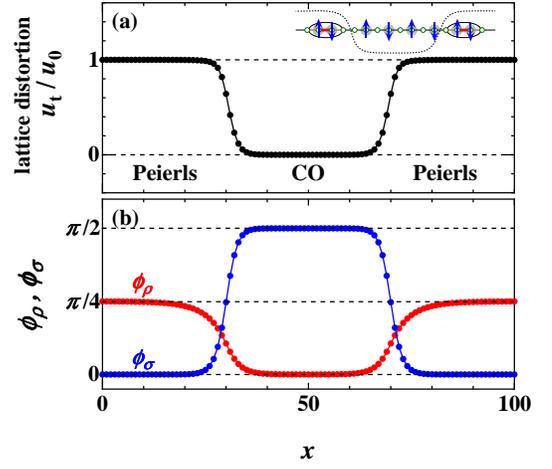}
\caption{
(Color online)
(a) The lattice distortion for the coexisting 
 CO and Peierls states 
(the schematic view is also shown)
and
(b) the corresponding  soliton-antisoliton formation 
   in the phase variables $\phi_\rho$ and $\phi_\sigma$.
We have set
 $g_{1/4}/g_\sigma=-1/4$, $K_{\mathrm{ph}}=1/(5g_\sigma)$, 
$K_\rho=0.1$, $K_\sigma=1$, and $v_\rho=v_\sigma=\sqrt{2} t$
with
$g_\sigma=2t$.
}
\label{fig:CO_domain}
\end{figure}

 We consider the case wherein both the Peierls state 
  and the CO state become locally stable. 
Even for the case in which
   the true ground state is given by the Peierls state, 
  it is actually expected that the Peierls state coexists with
   the metastable CO state without lattice distortion,
    in the photoexcited phase of (EDO-TTF)$_2$PF$_6$.

 Figure \ref{fig:CO_domain} shows
   the coexisting state between the Peierls state and 
the SDW+CO state, 
  created by the soliton-antisoliton formation
  connecting the different two states,
  which is obtained by solving Eq.\ (\ref{eq:soliton}).
The parameters are set as
 $g_{1/4}/g_\sigma=-0.25$ and $g_\sigma K_{\mathrm{ph}}=0.2$, 
 which correspond to the values on the boundary
(see Fig.\ \ref{fig:phase_diagram}).
The parameters for the kinetic part are chosen as
$K_\rho=0.1$, $K_\sigma=1$, and $v_\rho=v_\sigma=\sqrt{2} t$.
 There exists a domain of  the metastable CO state with 
   $(u_{t},\phi_\rho,\phi_\sigma) = (0,0, \pi/2)$   
     in the  Peierls state with 
   $(u_{t},\phi_\rho,\phi_\sigma) = (u_{t}^0,\pi/4, 0)$. 
Note that the boundary between the Peierls and 
     the CO  states  is created  by the formation of 
        solitons for charge, spin, and lattice distortion. 
 Since such a coexisting state is expected 
not only for the Peierls phase at finite temperature
   but also for the state after the photoinduced phase transition
  in (EDO-TTF)$_2$PF$_6$,
   \cite{Chollet2005}
  it can be expected that such a domain stays  
   as long as the thermal fluctuations 
     cannot go over the potential barrier given by Eq.\
     (\ref{eq:height_CO}).

\subsection{Purely electronic excitations: 
the antiadiabatic limit}

Now, we examine the excited states induced only by electronic
 excitations;
i.e., spin and charge soliton-antisoliton excitations, while 
   we assume that  the lattice distortion remains uniform in space. 
This situation could be related to the 
  electronic excitations due to the probe light in 
  reflectivity measurements.
Such a soliton excitation is also expected to play crucial roles
     in the photoinduced phase transition in (EDO-TTF)$_2$PF$_6$
      due to the ultrafast photoresponse within the order of pico
     seconds.
In general,\cite{Yonemitsu2006}
  the lattice should locally relax
   under friction to the minimum of the adiabatic
 potential,
  when  photons are absorbed by  electrons 
 at a site.

\begin{figure}[t]
\includegraphics[width=7.5cm]{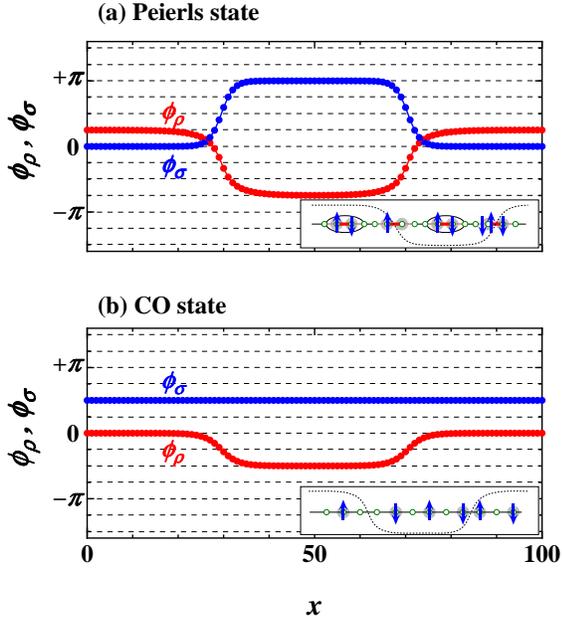}
\caption{
(Color online)
Soliton-antisoliton excitation 
of the charge and spin phase variables,
  with fixed several values of (a) $u_{t}=u_{t}^0$ 
  (corresponding to the Peierls state)
  and (b) $u_{t}=0$ (corresponding to the CO state).
The  soliton and/or antisoliton of $\phi_\rho$ and $\phi_\sigma$
  connects two neighboring minima of
 $V(u_{t},\phi_\rho,\phi_\sigma)$
 (see Fig.\ \ref{fig:potmin}).  
The low-energy excitation in the Peierls state  
  is a single-electron excitation, while that in the 
  CO state is a domain excitation.
The schematic views of the excitations for the respective cases
  are also shown. 
We have set $g_{1/4}/g_\sigma=-1/4$ and
$K_{\mathrm{ph}}=3/(16 g_\sigma)$ with $g_\sigma=2t$, 
 $K_\rho=0.1$, and $K_\sigma=1$.
}
\label{fig:soliton}
\end{figure}

To this end,
we calculate Eqs.\ (\ref{eq:phi_rho_opt}) and (\ref{eq:phi_sigma_opt}) 
 with the fixed uniform $u_{t}$,
  for the Peierls state ($u_{t}=u_{t}^0$)
   and the CO state ($u_{t}=0$). 
In the Peierls state corresponding to finite 
$u_{t}$ $(= u_{t}^0)$, 
 the soliton-antisoliton excitation is characterized as
\begin{eqnarray}
(\phi_\rho,\phi_\sigma)
= 
\left(\frac{\pi}{4},0\right)
 \to
\left(-\frac{3\pi}{4},\pi\right)
 \to
\left(\frac{\pi}{4},0\right) . 
\label{eq:soliton_Peierls}
\end{eqnarray}
The explicit profile is shown in Fig.\ \ref{fig:soliton}(a).
Thus, the soliton-antisoliton excitation in the Peierls state carries
  the topological charge $Q=\pm 1$ and the spin 
  $S_z=\pm 1/2$, \cite{Tsuchiizu2004}
which is nothing but a single-electron excitation.
The simple picture is the following:
A particle at the position of the soliton  moves away from  
    the place of the antisoliton  
     by breaking  the spin-singlet state.
Then, one finds   an emergence of 
   a local spin with a hole around  the  soliton kink 
   and that of    a local spin with an extra electron
    around  the  antisoliton kink. 
On the other hand, the soliton in the CO state 
is  given by [Fig.\ \ref{fig:soliton}(b)],
\begin{eqnarray}
(\phi_\rho,\phi_\sigma)
  =
  \left(0,\frac{\pi}{2}\right) 
  \rightarrow 
\left(-\frac{\pi}{2},\frac{\pi}{2}\right)
 \rightarrow 
\left(0,\frac{\pi}{2}\right) , 
\label{eq:soliton_CO}
\end{eqnarray}
which carries the topological charge 
   $Q=\pm 1/2$ and $S_z=0$.
A noticeable feature is seen from the fact that  
  only the charge excitation occurs.  
This is a typical domain excitation in the CO state.
The charge increases at the kink of the soliton 
 and the extra hole appears at the kink of the antisoliton.

\begin{figure}[t]
\includegraphics[width=6.5cm]{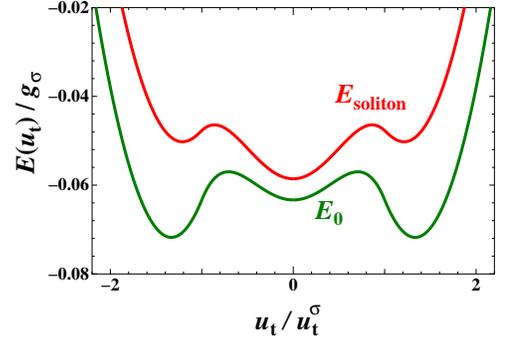}
\caption{
(Color online)
Ground-state ($E_0$) and soliton-antisoliton formation ($E_s$) energies 
 as a function of $u_{t}$, 
where we have set 
$g_{1/4}/g_\sigma=-1/4$ and $K_{\mathrm{ph}}=3/(16 g_\sigma)$ 
with $g_\sigma=2t$, $K_\rho=0.1$, and $K_\sigma=1$.
}
\label{fig:solitonenergy}
\end{figure}

As shown in Fig.\ \ref{fig:potmin}, the locking 
  location of the phase variables
  $\phi_\rho$ and $\phi_\sigma$ changes depending on the 
  strength of the lattice distortion  $u_{t}$. 
Even though the state with the intermediate
  values of $0<u_{t}<u_{t}^0$
  is not stable, we can consider
  the purely electronic soliton excitations 
  in such a virtual state and discuss the 
  ``$u_{t}$ dependence'' of the soliton excitation 
energy.
The $u_{t}$ dependence of the ground-state energy and the 
soliton-antisoliton energy is shown in Fig.\ \ref{fig:solitonenergy}.
Here, we assumed that the soliton and antisoliton do not form 
  a bound state (breather) and estimated 
  the soliton-antisoliton formation energy by doubling the
  soliton energy in  Fig.\ \ref{fig:soliton}.
 From   Fig.\ \ref{fig:solitonenergy},
 it is found that 
   the soliton formation energy in the CO state  costs less energy 
           compared with that in the Peierls state.
This is due to the fact that
   only the charge degree of freedom participates in 
   for the CO state.   
From this excitation energy analysis,
it can be found that both the Peierls state
  ($u_{t}=u_{t}^0$) and the
  CO state ($u_{t}=0$) have finite energy gaps 
  to the excited states; i.e., both states are insulators. 
However, in the CO state, the system could have a semiconducting 
   feature since the gap can become very small.
The magnitude of this energy gap depends on the 
  choice of parameters, but it
  is discussed by comparing the mean-field results 
  in Sec.\ \ref{sec:summary}.

There are many choices of creating solitons
in Fig.\ \ref{fig:soliton};
 however, 
the present estimation has been done for 
 the soliton, which has the minimum energy 
 for large  $u_{t}$ limit and is adiabatically connected to the 
  small $u_{t}$ case.
From this excitation energy analysis,
  it can be found that 
  the one-particle excitations
  in the Peierls state could be related 
  to the domain excitations in the CO state.
Such a connectivity would be relevant to the photoinduced phase 
transition mechanism, but this needs further investigation.

\section{Summary and Discussions}\label{sec:summary}

We have examined the exotic Peierls state observed in 
 (EDO-TTF)$_2$PF$_6$ and analyzed
the excitation from the Peierls ground state and also 
the excitation from the metastable CO state.
The metastable CO state around the CO state
    has been found 
    by considering  the modulation 
     of the elastic constants $\delta K_{P}$ 
 and $\delta K_{H}$
      and by considering 
  the moderate strength of the nearest neighbor repulsive interaction.  
The metastable state, which coexists with the spin-density wave, 
 exhibits soliton excitations, having an energy 
    much smaller than  that of the Peierls state. 
    The energy of the ground state and the excited state, which are obtained  
    as a function of $u_{t}$, exhibits a common feature 
     with the adiabatic potential in the electron-phonon coupled system in 
     one dimension.
     \cite{Koshino1998} 
Considering the correspondence between such a situation to
  the case of EDO-TTF, 
   the  present metastable state could be relevant to the state 
     of  the photoinduced phase.

Here, we discuss the experimental findings 
  on the reflectivity in the photoinduced phase 
 of (EDO-TTF)$_2$PF$_6$ based on the present 
  analysis.
 From the photoconductivity measurement, the photoinduced phase
  exhibits  a metallic behavior, although it is 
  not the same as that of  high-temperature metallic  state,
  as shown by the reflectivity data.\cite{Onda2005}
In the present analysis, we obtained the metastable state
 without lattice distortion, which  is not the normal state.
 Actually, in such a state,
    the translational invariance is broken due to the 
  presence of the CO state, which gives a small soliton excitation gap.
Here, we estimate the magnitude of the soliton gaps. 
From the mean-field approach,  \cite{Suzumura1997}
the commensurability potential is estimated as 
$C_0 \cos 4\phi_\rho$,
where the amplitude is given by
$C_0 \approx -0.03t$ for  $(U/t,V/t)=(4,2)$.
On the other hand,
 the amplitude of the $\cos 2\phi_\sigma$ potential
 \cite{Tomio2000JPSJ}
  is about 
  $0.15 t \sim 0.3 t$, which are estimated from 
  $(V/t,V_2/t)\approx (1.4,\, 1.05) \sim (2,\, 0)$.
Thus, we find $g_{1/4}/g_\sigma \approx  -0.1 \sim -0.2$ 
  and our choice of the parameter $g_{1/4}/g_\sigma=-1/4$ 
 in Fig.\ \ref{fig:solitonenergy}
 is not
  far from the mean-field results.
Actually,  we obtain 
$g_\sigma/(2\pi^2 a^2)=-4g_{1/4}/(2\pi^2a^2)
  = 0.12t $ from 
 the mean-field result,  $C_0(V=2t)\approx -0.03t$. 
Here, we note the soliton gap  at $u_{t}=0$, which  
  is of the order of 
 $\delta E(0) = [E_{\mathrm{soliton}}(0)-E_0(0)]
  \approx 0.01 t $,
and then 
 $\delta E(0) \approx  3$ meV. 
On the other hand, the soliton gap from the Peierls ground state is given by  
 $\delta E(u_{t}^{\mathrm{opt}}) 
= [E_{\mathrm{soliton}}(u_{t}^{\mathrm{opt}})
  -E_0(u_{t}^{\mathrm{opt}})]
  \approx 0.045 t \approx 14 $ meV,
   which is much larger than that of the metastable state.

Finally,  we comment on the quantum fluctuation on the lattice distortion, 
which is not treated in the present calculation. 
Such an effect appears when the term $\sum_{j}P_j^2/2M$ is 
taken into account for 
  Eq.\ (\ref{eq:Hph}) with  $[u_j,P_{j'}]= \delta_{j,j'}$. 
The present calculation corresponds to the limit of a large mass $M$. 
If we naively consider a renormalization 
 by   a  Debye$-$Waller factor, \cite{Ziman}
 the effective coupling constant 
   $g_{\mathrm{ph}}$ is reduced due to the quantum fluctuation. 
  Thus, this fluctuation  makes the normalized  constant $K_{\mathrm{ph}}$ 
  large and  the region for CO in Fig.\ \ref{fig:phase_diagram}
   is extended.  
  For  the limit of the light mass, which corresponds to 
     the antiadiabatic case,  
   the averaged lattice distortion vanishes but 
   the crossover from the spin-density-wave state 
     to the spin singlet is expected with an increase in the 
  electron-phonon coupling 
      constant $g_{\rm ph} (> \sqrt{K_{\rm ph} U})$. 
      \cite{Yonemitsu1996PRB,Tsuchiizu2005} 
  Thus,  the metastable state, which coexists with 
  the CO and spin-density-wave states due to the electronic correlations,
   is expected to give rise to the conduction by forming the solitary 
    charge excitation if the quantum fluctuations of both the electron 
     and  the phonon  remain moderately small.

\acknowledgments

The authors are grateful to H.\ Yamochi, G.\ Saito, A.\ Ota, Y.\ Nakano,
  S.\ Koshihara, and K.\ Onda,
for valuable discussions on the experimental findings.
They also thank Y.\ Omori, H.\ Seo, H.\ Fukuyama, K.\ Yonemitsu, 
J.\ F.\ Halet,  and T.\ Giamarchi, for
  useful discussions on the theoretical aspects. 
The authors are also indebted to T.\ Mori and T.\ Kawamoto for the 
 helpful discussions on the extended H\"uckel method.
 This work was supported by a Grant-in-Aid for
Scientific Research on Priority Areas of Molecular Conductors
(Grant No.\ 15073213) from 
 the Ministry of Education, Science, Sports, and Culture of Japan.


\begin{thebibliography}{}
\bibitem{Chollet2005}
M.\ Chollet, L.\ Guerin, N.\ Uchida, 
S.\ Fukaya, H.\ Shimoda, T.\ Ishikawa,
K.\ Matsuda, T.\ Hasegawa, A.\ Ota,
H.\ Yamochi, G.\ Saito, R.\ Tazaki, 
S.\ Adachi, and S.\ Koshihara,
Science \textbf{307}, 86 (2005).
\bibitem{Onda2005}
K.\ Onda, T.\ Ishikawa, M.\ Chollet,
X.\ Shao, H.\ Yamochi, G.\ Saito, and S.\ Koshihara,
J.\ Phys:\ Conf.\ Ser.\ \textbf{21}, 216 (2005).
\bibitem{Koshihara2006}
S.\ Koshihara and S.\ Adachi, 
J.\ Phys.\ Soc.\ Jpn.\ \textbf{75}, 011005 (2006).
\bibitem{Ota2002}
A.\ Ota, H.\ Yamochi, and G.\ Saito,
J.\ Mater.\ Chem.\ \textbf{12}, 2600 (2002);
A.\ Ota, H.\ Yamochi, and G.\ Saito,
Synth.\ Met.\ \textbf{133-134}, 463 (2003).
\bibitem{Drozdova2004}
O.\ Drozdova, K.\ Yakushi, 
A.\ Ota, H.\ Yamochi, and G.\ Saito,
Synth.\ Met.\  \textbf{133-134}, 277 (2003);
O.\ Drozdova, K.\ Yakushi, K.\ Yamamoto, 
A.\ Ota, H.\ Yamochi, G.\ Saito,
H.\ Tashiro, and D.\ B.\ Tanner,
Phys.\ Rev.\ B  \textbf{70}, 075107 (2004).
\bibitem{Tokura2006}
Y.\ Tokura, 
J.\ Phys.\ Soc.\ Jpn.\ \textbf{75}, 011001 (2006).
\bibitem{Ung}
K.\ C.\ Ung, S.\ Mazumdar, and D.\ Toussaint,
Phys.\ Rev.\ Lett.\ \textbf{73}, 2603 (1994).
\bibitem{Clay}
R.\ T.\ Clay, S.\ Mazumdar, and D.\ K.\ Campbell,
Phys.\ Rev.\ B \textbf{67}, 115121 (2003).
\bibitem{Kuwabara2003}
M.\ Kuwabara, H.\ Seo, and M.\ Ogata,
J.\ Phys.\ Soc.\ Jpn.\ \textbf{72}, 225 (2003).
\bibitem{Seo2007JPSJ}
H.\ Seo, Y.\ Motome, and T.\ Kato,
J.\ Phys.\ Soc.\ Jpn.\ \textbf{76}, 013707 (2007).
\bibitem{Omori2007}
Y.\ Omori, M.\ Tsuchiizu, and Y.\ Suzumura,
J.\ Phys.\ Soc.\ Jpn.\ \textbf{76}, 114709 (2007).
\bibitem{Mori_extdh}
T.\ Mori, A.\ Kobayashi, Y.\ Sasaki, H.\ Kobayashi, 
G.\ Saito, and H.\ Inokuchi, 
Bull.\ Chem.\ Soc.\ Jpn.\ \textbf{57}, 627 (1984); 
T.\ Mori, Ph.D.\ thesis, University of Tokyo, 1985.
\bibitem{Sugiura2003}
M.\ Sugiura, and Y.\ Suzumura,
J.\ Phys.\ Soc.\ Jpn.\ \textbf{72}, 1458 (2003);
M.\ Sugiura, M.\ Tsuchiizu, and Y.\ Suzumura,
\textit{ibid.} \textbf{74}, 983 (2005).
\bibitem{Giamarchi_book}
For a review, see T.\ Giamarchi,
\textit{Quantum Physics in One Dimension}
(Oxford University Press, New York, 2004).
\bibitem{Fukuyama_review}
H.\ Fukuyama and H.\ Takayama,
in \textit{Electronic Properties 
  of Inorganic Quasi-One-Dimensional Compounds}, 
  ed. P.\ Monceau 
(Reidel, Dordrecht, 1985), p.\ 41.
\bibitem{Tsuchiizu2001}
M.\ Tsuchiizu, H.\ Yoshioka, and Y.\ Suzumura,
J.\ Phys.\ Soc.\ Jpn.\ \textbf{70}, 1460 (2001).
\bibitem{Hara1983}
J.\ Hara and H.\ Fukuyama,
J.\ Phys.\ Soc.\ Jpn.\ \textbf{52}, 2128 (1983).
\bibitem{Takayama1980}
H.\ Takayama, Y.\ R.\ Lin-Liu, and K.\ Maki,
Phys.\ Rev.\ B \textbf{21}, 2388 (1980).
\bibitem{Yonemitsu2006}
K.\ Yonemitsu and K.\ Nasu,
J.\ Phys.\ Soc.\ Jpn.\ \textbf{75}, 011008 (2006).
\bibitem{Tsuchiizu2004}
M.\ Tsuchiizu and A.\ Furusaki,
Phys.\ Rev.\ B \textbf{69}, 035103 (2004).
\bibitem{Koshino1998}
K.\ Koshino and T.\ Ogawa,
J.\ Phys.\ Soc.\ Jpn.\ \textbf{67}, 2174 (1998).
\bibitem{Suzumura1997}
Y.\ Suzumura,
J.\ Phys.\ Soc.\ Jpn.\ \textbf{66}, 3244 (1997).
\bibitem{Tomio2000JPSJ} 
Y.\ Tomio and Y.\ Suzumura,
J.\ Phys.\ Soc.\ Jpn.\ \textbf{69}, 796 (2000).
\bibitem{Ziman} 
J.\ M.\ Ziman,
 \textit{Principles of the Theory of Solids}
  (Cambridge University Press, Cambridge, 1965), p.\ 60.
\bibitem{Yonemitsu1996PRB}
K.\ Yonemitsu and M.\ Imada,
Phys.\ Rev.\ B \textbf{54}, 2410 (1996).
\bibitem{Tsuchiizu2005}
M.\ Tsuchiizu, M.\ Sugiura,  and Y.\ Suzumura,
Physica B (Amsterdam) \textbf{358}, 42 (2005).




\end{thebibliography}
\end{document}